\documentclass[useAMS,usenatbib,fleqn,twocolumn]{aastex631}
\usepackage{mathtools}
\usepackage{amssymb,amsmath}
\usepackage[T1]{fontenc}

\begin{document}

\shorttitle{Observing Frame Dragging}
\title{Observational Signatures of Frame Dragging in Strong Gravity}

\shortauthors{Ricarte et al.}
\correspondingauthor{Angelo Ricarte}
\email{angelo.ricarte@cfa.harvard.edu}

\author{Angelo Ricarte}
\affiliation{Center for Astrophysics | Harvard \& Smithsonian, 60 Garden Street, Cambridge, MA 02138, USA}
\affiliation{Black Hole Initiative, 20 Garden Street, Cambridge, MA 02138, USA}

\author{Daniel~C.~M.~Palumbo}
\affiliation{Center for Astrophysics | Harvard \& Smithsonian, 60 Garden Street, Cambridge, MA 02138, USA}
\affiliation{Black Hole Initiative, 20 Garden Street, Cambridge, MA 02138, USA}

\author{Ramesh Narayan}
\affiliation{Center for Astrophysics | Harvard \& Smithsonian, 60 Garden Street, Cambridge, MA 02138, USA}
\affiliation{Black Hole Initiative, 20 Garden Street, Cambridge, MA 02138, USA}

\author{Freek Roelofs}
\affiliation{Center for Astrophysics | Harvard \& Smithsonian, 60 Garden Street, Cambridge, MA 02138, USA}
\affiliation{Black Hole Initiative, 20 Garden Street, Cambridge, MA 02138, USA}

\author{Razieh Emami}
\affiliation{Center for Astrophysics | Harvard \& Smithsonian, 60 Garden Street, Cambridge, MA 02138, USA}

\date{\today}

\begin{abstract}
Objects orbiting in the presence of a rotating massive body experience a gravitomagnetic frame-dragging effect, known as the Lense-Thirring effect, that has been experimentally confirmed in the weak-field limit.  In the strong-field limit, near the horizon of a rotating black hole, frame dragging becomes so extreme that all objects must co-rotate with the black hole's angular momentum.  In this work, we perform general relativistic numerical simulations to identify observable signatures of frame dragging in the strong-field limit that appear when infalling gas is forced to flip its direction of rotation as it is being accreted.  In total intensity images, infalling streams exhibit ``S''-shaped features due to the switch in the tangential velocity.  In linear polarization, a flip in the handedness of spatially resolved polarization ticks as a function of radius encodes a transition in the magnetic field geometry that occurs due to magnetic flux freezing in the dragged plasma.  Using a network of telescopes around the world, the Event Horizon Telescope collaboration has demonstrated that it is now possible to directly image black holes on event horizon scales.  We show that the phenomena described in this work would be accessible to the next-generation Event Horizon Telescope (ngEHT) and extensions of the array into space, which would produce spatially resolved images on event horizon scales with higher spatial resolution and dynamic range.
\end{abstract}


\section{Introduction}
\label{sec:introduction}

In general relativity, inertial frames orbiting a massive rotating body are dragged by a gravitomagnetic effect known as the Lense-Thirring effect \citep{Lense&Thirring1918}.  This phenomenon has been experimentally verified in the weak-field regime by satellites orbiting the Earth \citep{Ciufolini&Pavlis2004,Everitt+2011} and possibly detected in a white dwarf-pulsar binary system \citep{VenkatramanKrishnan+2020}.  In both cases, the effect is weak.  Near the horizon of a black hole (BH), frame dragging becomes more extreme.  Inside a region called the ergosphere, all bodies are forced to co-rotate with the BH (in Boyer-Lindquist coordinates).

The Event Horizon Telescope (EHT) has produced the first spatially resolved BH images, ushering in a new era of spatially resolved BH astrophysics in the strong field regime \citep{EHT1,EHT2,EHT3,EHT4,EHT5,EHT6,EHT7,EHT8,EHT_SgrA_I,EHT_SgrA_II,EHT_SgrA_III,EHT_SgrA_IV,EHT_SgrA_V,EHT_SgrA_VI}.  These images impose direct constraints on BH accretion flow models, favoring those with unequal ion and electron temperatures and dynamically important magnetic fields \citep{EHT5,EHT8}.  The next-generation EHT (ngEHT) will deliver higher spatial resolution and a much greater dynamic range, enabling more detailed image reconstructions and even movies that will simultaneously resolve the inner accretion flow and the jet in the next decade \citep{Doeleman+2019,Raymond+2021}.  In the more distant future, expanding the array into space would allow for much greater spatial resolution as well as more rapid imaging cadences \citep{Roelofs+2019,Palumbo+2019,Roelofs+2021}.  With access to higher spatial resolution, one could access more targets as well as sharper image features such as the photon ring \citep[e.g.,][]{Johnson+2020}.

In this letter, we identify two direct observational signatures of frame dragging unique to retrograde accretion flows, those where the disk and BH angular momentum vectors are anti-aligned.  Because of the existence of the ergosphere in the Kerr metric, there must exist a radius at which the tangential velocity changes sign.  As we shall show, this transition can impart signatures in total intensity images and linear polarization maps that could be imaged by extensions to the EHT.  

In \autoref{sec:methodology}, we describe how we generate simulated images of both prograde and retrograde accretion flows and devise metrics in both total intensity and polarization that select retrogrades.  In \autoref{sec:results}, we show that metrics to select for changes in ``handedness'' of infalling streams and linear polarization morphology can select for retrograde systems.  We conclude in \autoref{sec:conclusion} by summarizing our results and discussing implications for telescope designs that can observe these features. 

\section{Methodology}
\label{sec:methodology}

\subsection{GRMHD}
\label{sec:grmhd}

We use as our starting point 18 different GRMHD simulations performed with the code {\sc koral} \citep{Sadowski+2013,Sadowski+2014}.  These simulations include 9 different dimensionless spin values $a_\bullet \in \{0,\pm 0.3,\pm 0.5,\pm 0.7,\pm 0.9\}$ and 2 different magnetic field states, both ``magnetically arrested disk'' (MAD) models and ``standard and normal evolution'' (SANE) models.  MAD models are characterized by dynamically important magnetic fields that can lead to flux eruption events \citep{Bisnovatyi-Kogan&Ruzmaikin1974,Igumenshchev+2003,Narayan+2003}, while the magnetic fields in SANE models are turbulent and dynamically unimportant \citep{Narayan+2012,Sadowski+2013}.  A model is ``prograde'' if the disk's angular momentum vector is aligned with that of the BH, denoted with $a_\bullet>0$, and ``retrograde'' if the two vectors are anti-aligned, denoted with $a_\bullet<0$.  The MAD models were run to $t \approx 10^5 \ GM_\bullet/c^3$ and were presented in \citep{Narayan+2022}.  The SANE models introduced in this work use the same basic setup as the MAD simulations, except that the initial magnetic field configuration was a set of quadrupolar poloidal field loops instead of a single dipolar loop.  This change inhibited the growth of the poloidal magnetic field threading the black hole and ensured that the accretion flow remained in the SANE state until the end of the simulation at $t=3 \times 10^4 \ GM_\bullet/c^3$.

\begin{figure*}
  \centering
  \begin{tabular}{c}
  \includegraphics[width=\textwidth]{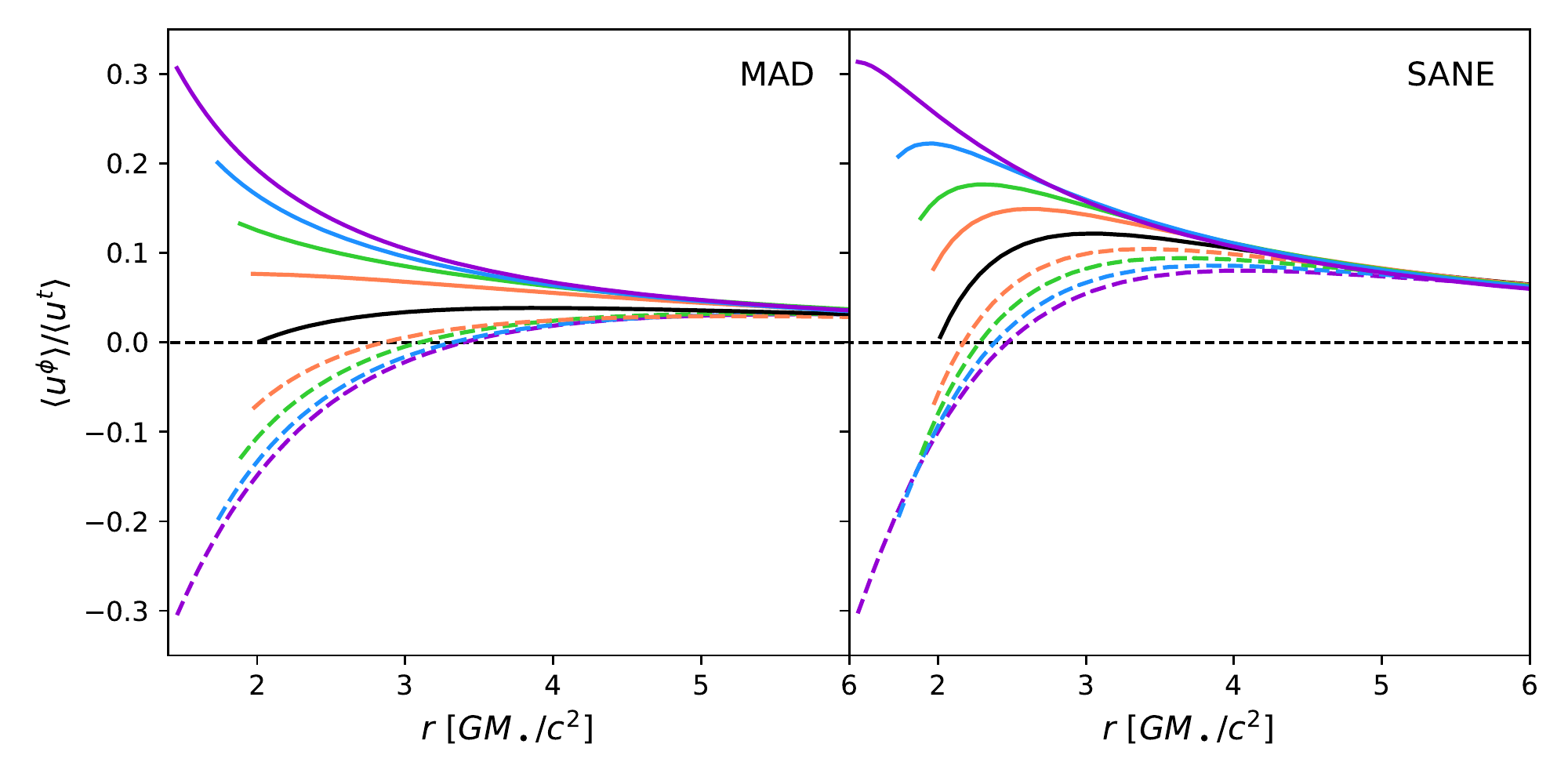} \\
  \includegraphics[width=\textwidth]{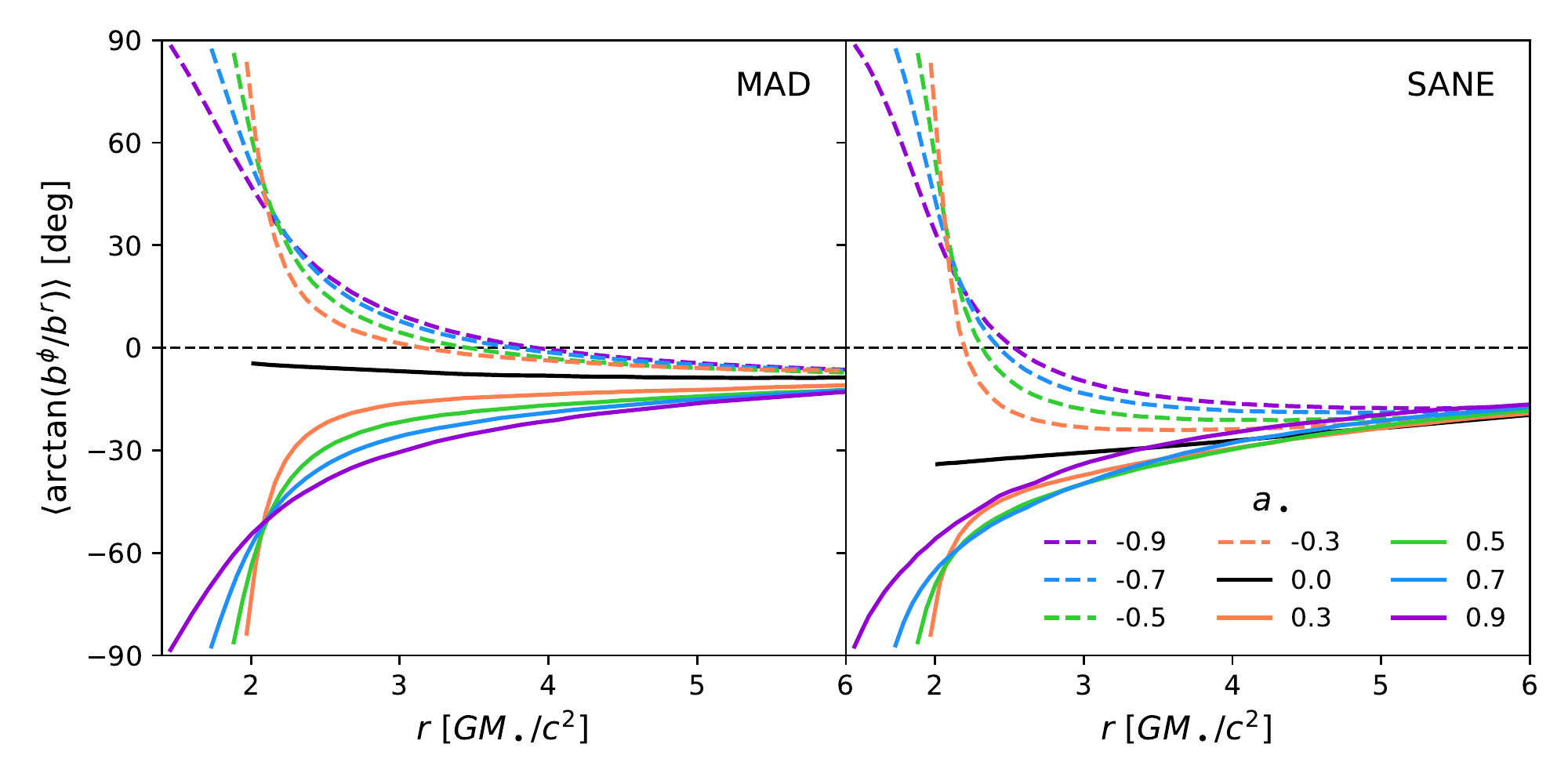}
  \end{tabular}
  \caption{Tangential velocity (top) and angle of the magnetic field in the midplane (bottom) in the midplane of our GRMHD simulations, averaged over both azimuth and time.  Quantities are computed in Boyer-Lindquist coordinates in the lab frame.  Both plotted quantities change sign as a function of radius {\it if and only if} the system is retrograde (dashed lines), reflecting a transition in the dynamics of the accretion flow.}  \label{fig:fluid}
\end{figure*}

In \autoref{fig:fluid}, we plot azimuthal- and time-averages of the tangential velocity\footnote{We have corrected an error in Figure 7 of \citet{Narayan+2022} which led to an incorrect additional flip in the tangential velocity within the ergosphere.} ($u^\phi/u^t$) and angle of the magnetic field in the midplane ($\arctan(b^\phi/b^r)$) in Boyer-Lindquist coordinates for the different simulations.  Different colors encode different values of $|a_\bullet|$, and retrograde models are plotted with dashed lines.  Both of these quantities change sign {\it if and only if} the simulation is retrograde.  The transition in the fluid dynamics is forced to occur due to the existence of the ergosphere, which forces a retrograde flow to change rotational direction at some radius.  Then, due to flux freezing in ideal GRMHD, a similar transition occurs in the geometry of the magnetic field.  As $|a_\bullet|$ increases, the region that is forced to co-rotate with the BH increases, and the radius at which $u^\phi/u^t$ and $\arctan(b^\phi/b^r)$ change sign moves outward.

\subsection{GRRT}
\label{sec:grrt}

From these GRMHD snapshots, we produce model images using the polarized general relativistic radiative transfer (GRRT) code {\sc ipole} \citep{Moscibrodzka&Gammie2018}, following image generation techniques detailed in \autoref{sec:image_generation}.  We explore 4 different values of the parameter $R_\mathrm{high}\in \{1,10,40,160\}$, which sets the asymptotic ratio of ion to electron temperatures in regions where gas pressure dominates over magnetic pressure \citep{Moscibrodzka+2016}.  We consider only thermal electron distribution functions when evaluating the radiative transfer coefficients \citep{Dexter2016}.

For each combination of magnetic field state, spin, and $R_\mathrm{high}$, we image 201 snapshots evenly spaced between times $10^4 \ GM_\bullet/c^3$ (enough time for the accretion disk to reach steady state), and the end of the simulation.  All models are scaled to the appropriate mass, distance and flux of M87*\footnote{We set the mass and distance to $6.2\times 10^9 \ M_\odot$ and 16.9 Mpc respectively \citep{Gebhardt+2011}, and set the average flux to 0.5 Jy \citep{EHT5}.}, the primary EHT and ngEHT observing target.  We assume a viewing angle of $17^\circ$ with respect to the approaching jet \citep{Walker+2018}, or $163^\circ$ for our prograde models, to ensure that the BH rotates clockwise on the sky.  Sagittarius A* could equally well exhibit the observational signatures described in this work, but the existence of a scattering screen in the millimeter \citep[e.g.,][]{Issaoun+2019} and more rapid time variability makes M87* an easier target to observe.  Images are computed at 86, 228, 345, and 690 GHz, but we focus our presentation on 228 GHz, the central observing frequency of 2017 EHT observations \citep{EHT1}.

\subsection{Metrics for Spiral Handedness in Total Intensity}
\label{sec:handedness_definition}

Model images exhibit streams of gas plunging into the BH from the accretion disk that can potentially be resolved by the ngEHT.  Of particular interest to our study is the apparent handedness of these streams, clockwise or counterclockwise, which encodes the azimuthal velocity direction.  Retrograde images can exhibit streams that unambiguously switch handedness as a function of radius by eye.  To characterize these flips in handedness in a quantitative and computationally tractable manner, we devise two different metrics for the spiral handedness of a given region.

\subsubsection{Fourier Decomposition Method}

In one method, we perform a Fourier decomposition of each model image into a series of logarithmic spirals.  In brief, we calculate complex Fourier coefficients $A(p,m)$ for $m\in \{0,1,2,...,20\}$ and $p \in [-100,100]$, where $m$ corresponds to the number of spiral arms in a mode, and $p$ is related to the pitch angle of a spiral arm.  Modes with $p>0$ correspond to clockwise modes, and modes with $p<0$ correspond to counter-clockwise modes, allowing us to determine the relative handedness of features in a given region in the image.  The full details of this Fourier decomposition are described in \autoref{sec:fourier_decomposition}.

\begin{figure*}
  \centering
  \includegraphics[width=\textwidth]{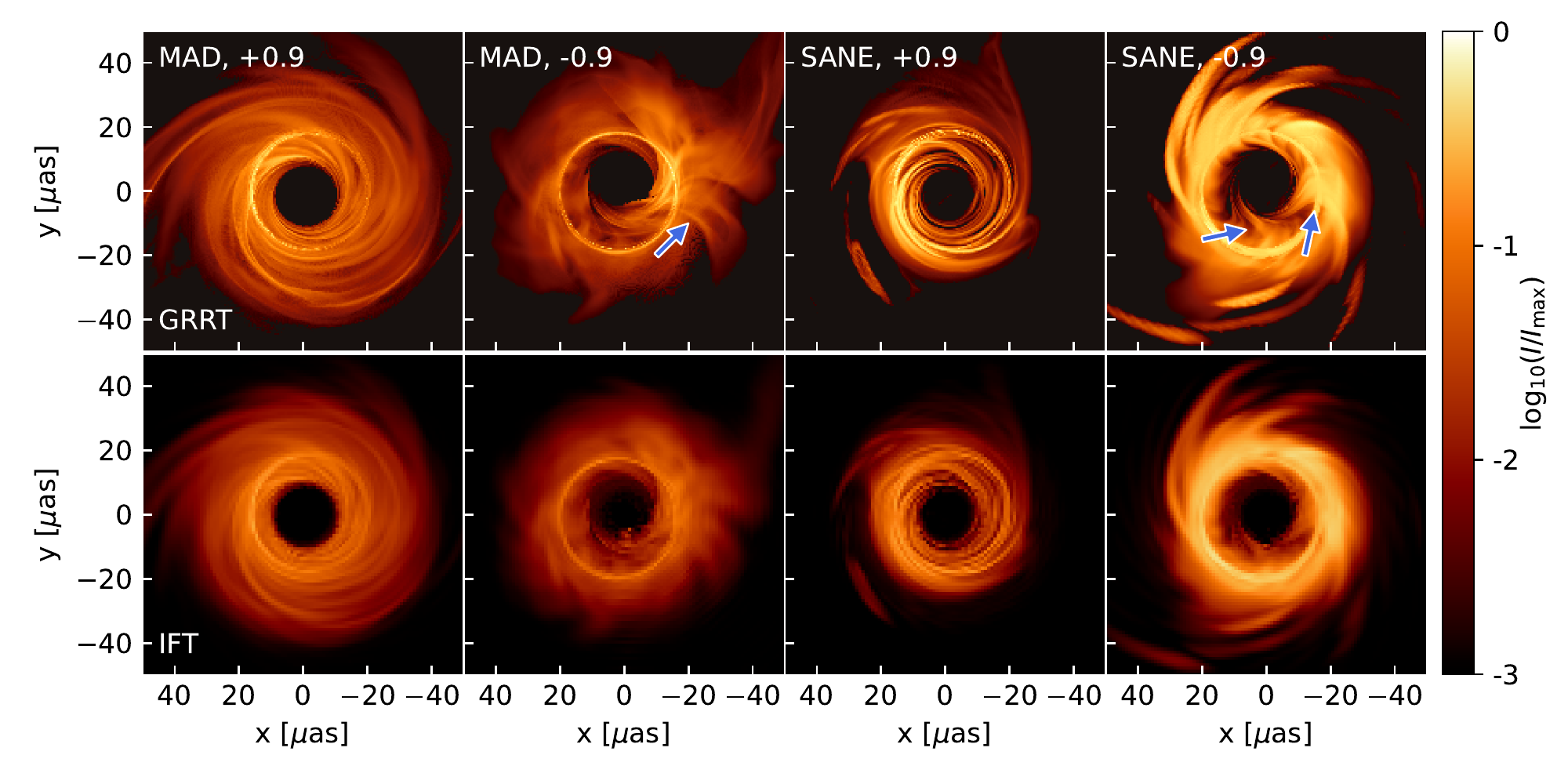}
  \caption{Four example images from our image library at 228 GHz (top) along with their inverse Fourier transforms (IFTs) following decompositions into logarithmic spirals (bottom).  All images have $R_\mathrm{high}=40$, with magnetic field states and spins as written at the top left of each column.  In each column, the value of $I_\mathrm{max}$ used to set the color scale is the maximum intensity in the GRRT image.  The logarithmic spiral decompositions with $m \leq 20$ do an excellent job at recovering even detailed image structure.  In the retrograde images, some streams flip handedness as a function of radius, forming ``S''-shaped patterns, some of which are marked with blue arrows.} \label{fig:examples}
\end{figure*}

In \autoref{fig:examples}, we plot a few representative snapshots from our model library, all of which have $R_\mathrm{high}=40$ and $|a_\bullet|=0.9$, corresponding to a prograde MAD, a retrograde MAD, a prograde SANE, and a retrograde SANE.  The top row displays the true image from our GRRT calculations, and the bottom row displays the inverse Fourier transform (IFT) following our Fourier decomposition technique.  Our logarithmic spiral decompositions with $m \leq 20$ perform exceptionally well at capturing even detailed structure within these images.  In the prograde images, spiral arms maintain the same handedness at all radii.  In contrast, some spiral arms in the retrograde images flip handedness at some radius, forming ``S''-shaped streams, some of which we mark with blue arrows.  In the retrograde SANE, this transition is unambiguous and occurs interior to the photon ring.  The retrograde MAD example contains both clockwise and counterclockwise features outside the photon ring, partially obfuscating this signature.  We find that this is due to the superposition of jet emission atop emission from the midplane where the flip in handedness originates.

With Fourier coefficients in hand, we devise a metric to summarize the handedness of a given region via

\begin{equation}
    H_f \equiv \left( \sum_{m=1}^{20} \int_0^{100} |A(p,m)| dp \right) \bigg/ \left( \sum_{m=1}^{20} \int_{-100}^0 |A(p,m)| dp \right).
    \label{eqn:handedness_fourier}
\end{equation}

\noindent That is, we stack the amplitudes corresponding to clockwise and counter clockwise modes respectively, then take the ratio.  As constructed, $H_f>1$ if there is more power in clockwise spirals and $H_f<1$ if there is more power in counterclockwise spirals.  Note that when computing $H_f$ we ignore $m=0$, which is symmetric by construction.

\subsubsection{Image Gradient Method}

To test the robustness of this metric, we also apply another metric based on gradients of the image.  Across each pixel, we compute the local handedness

\begin{equation}
    h_g = \mathrm{sign}\bigg(\frac{\partial f}{\partial \rho} \bigg/ \frac{\partial f}{\partial \varphi} \bigg)
\end{equation}

\noindent where $f$ is the distribution for which we are computing the handedness, such as the intensity, and $\rho$ and $\varphi$ are polar coordinates in the image.  Here, $h_g$ is $+1$ for locally clockwise spirals and $-1$ for locally counterclockwise spirals.  Then, we assign a single value of the handedness to a given region via 

\begin{equation}
    H_\mathrm{g} = \iint h_g f dx dy \bigg/ \iint f dx dy .
    \label{eqn:handedness_gradient}
\end{equation}

\noindent where the double integral represents a sum across a given area of the image.  In practice, this metric is more intuitive than the Fourier-based metric, but due to its local definition, it is likely more sensitive to imperfections in the data, such as image reconstruction artifacts.

\subsection{A Metric for Handedness of Linear Polarization}
\label{sec:beta2_definition}

The EHT observes synchrotron emission, whose linear polarization encodes the underlying magnetic field geometry.  \citet{Palumbo+2020} found that the ``twistiness'' of linear polarization ticks discriminated between MAD and SANE models of M87*.  In an annulus with inner and outer radii $\rho_\mathrm{min}$ and $\rho_\mathrm{max}$ respectively, the twistiness of linear polarization ticks can be quantified via the metric $\beta_2$, given by

\begin{align}
    \beta_2 &=\dfrac{1}{I_{\rm tot}} \int_{\rho_\mathrm{min}}^{\rho_\mathrm{max}} \int_0^{2 \pi} P(\rho, \varphi) \, e^{- 2i \varphi} \; \rho \mathop{d\varphi}  \mathop{d\rho},
\end{align}

\noindent with

\begin{align}
    I_{\rm tot} &= \int_{\rho_\mathrm{min}}^{\rho_\mathrm{max}} \int_0^{2 \pi} I(\rho, \varphi) \; \rho \mathop{d\varphi} \mathop{d\rho}.
\end{align}

\noindent This complex quantity is the rotationally symmetric component of a full Fourier decomposition of the linear polarization ticks.  The amplitude $|\beta_2|$ is the strength of this mode, and $\angle \beta_2$ in the complex plane encodes the pitch angle of the ticks.  Clockwise EVPA patterns have $\mathrm{Im}(\beta_2)>0$, while counterclockwise EVPA patterns have $\mathrm{Im}(\beta_2)<0$.  Hence, we can search for flips in the handedness of polarization ticks by searching for flips in the sign of $\mathrm{Im}(\beta_2)$ with radius.

\section{Results}
\label{sec:results}

\subsection{Handedness of Spirals in Total Intensity}

\begin{figure*}
  \centering
  \includegraphics[width=\textwidth]{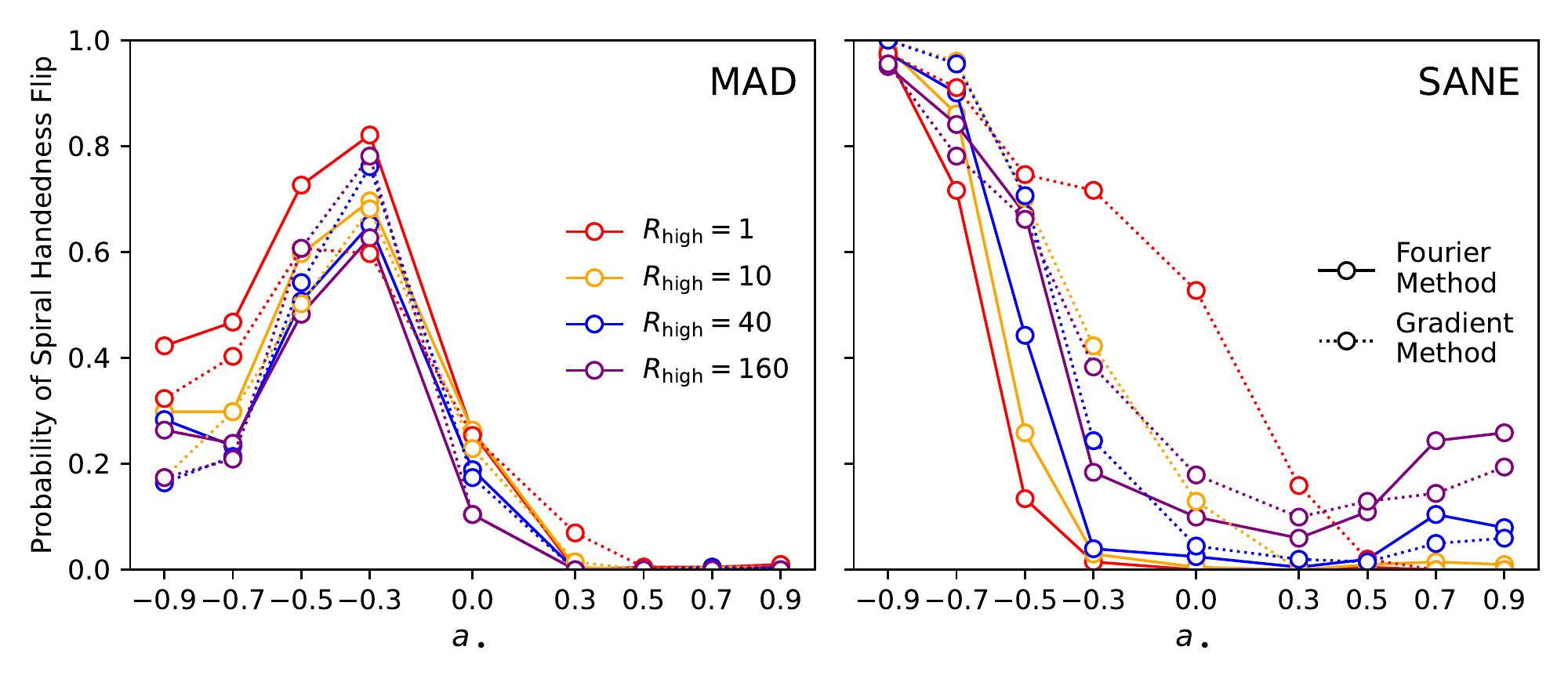}
  \caption{Probability of a flip in handedness of spiral features occurring between the interior and exterior regions, separated by the photon ring in our 228 GHz images.  Different values of $R_\mathrm{high}$ are shown in different colors.  Probabilities computed with the Fourier method (\autoref{eqn:handedness_fourier}) are plotted with solid lines, while probabilities computed with the gradient method  (\autoref{eqn:handedness_gradient}) are plotted with dotted lines.  Requiring a few independent high-resolution observations, this metric can accurately identify retrograde systems.  Note that our images have not been blurred for this analysis, since high spatial resolution of futuristic arrays is required to resolve these streams. \label{fig:handednessFlipProbabilities}}
\end{figure*}

Using both the Fourier method (\autoref{eqn:handedness_fourier}) and the gradient method (\autoref{eqn:handedness_gradient}), we compute the handedness in two annular regions for all images in our library, searching for a flip as a function of radius.  The interior and exterior annuli are defined by $\rho \in [1 \; \mu\mathrm{as},16\; \mu\mathrm{as}]$ and $\rho \in [31 \; \mu\mathrm{as},50 \; \mu\mathrm{as}]$ respectively, deliberately avoiding the photon ring.  In \autoref{fig:handednessFlipProbabilities}, we plot the probability that a given snapshot at 228 GHz exhibits a flip in handedness between these two regions as a function of spin.  Different colors encode different values of $R_\mathrm{high}$, and different linestyles correspond to the different methods.  As expected, the probability of there being a flip in handedness is much higher for retrograde than for prograde accretion flows.  Retrograde SANEs are more likely to exhibit this signature than retrograde MADs.  Unexpectedly, the signature is more difficult to detect for retrograde MADs with larger $|a_\bullet|$, regardless of the handedness metric chosen.  Although the signature exists for all retrograde spins in the GRMHD fluid variables, the transition moves to larger radii, where the intensity decreases (see \autoref{fig:fluid}).  In addition, the jet power in these models increases with $|a_\bullet|$ \citep{Narayan+2022}, and increased jet emission may be obfuscating this signature.  We find that the flip in handedness is more likely to exist in our 86 GHz images for larger values of $|a_\bullet|$, although the longer wavelength makes high-resolution imaging more difficult.

\subsection{Linear Polarization Twist as a Function of Radius}

Similarly, we compute $\angle \beta_2$ as a function of image radius to seek flips in handedness that would be observed in linear polarization.  Recall that $\mathrm{sign}(\mathrm{Im}(\beta_2))$ encodes the handedness of linear polarization ticks.  In the absence of Faraday and general relativistic effects, the handedness of these ticks directly reflects the projected geometry of the magnetic field.  To identify a flip in the handedness of linear polarization ticks, we first blur images with a Gaussian kernel with a full width at half maximum of $20 \ \mu\mathrm{as}$, which helps account for beam depolarization and scrambling.  Then, we consider an image to have an observable flip in the handedness of linear polarization ticks if there exist two annuli with (i) $|\beta_2|>10^{-2}$, (ii) intensities no less than a factor of 1000 less than the annulus with the peak intensity, and (iii) opposite signs of $\mathrm{Im}(\beta_2)$.  

\begin{figure*}
  \centering
  \includegraphics[width=\textwidth]{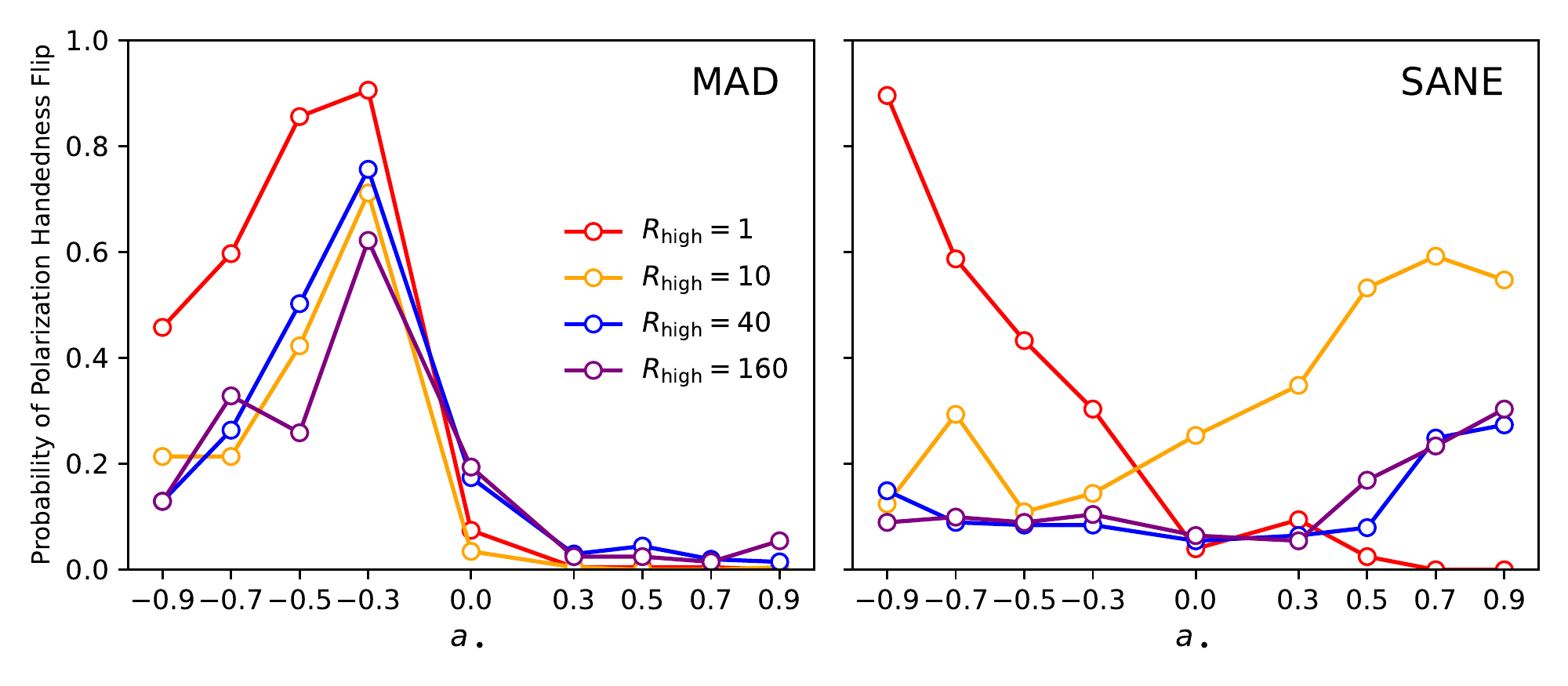}
  \caption{Probability of there existing an observable flip in the sign of $\angle\beta_2$ as a function of radius (a sign flip in its imaginary component) in our 228 GHz images.  Images have first been convolved with a Gaussian with a FWHM of 20 $\mu\mathrm{as}$.  For a flip to be detectable, we demand that its $|\beta_2|>10^{-2}$ and its intensity is no less than a factor of $10^3$ less than the annulus with the peak intensity.  This metric preferentially selects for retrograde MADs.  SANEs are more likely to fail the $|\beta_2|>10^{-2}$ condition due to their higher Faraday rotation depths.  Some prograde SANEs are erroneously selected by this metric because their overall $\angle \beta_2 \approx 0$, but such models could be ruled out with additional polarimetric information \citep{EHT8}. \label{fig:beta2_handedness_flip}}
\end{figure*}

In \autoref{fig:beta2_handedness_flip}, we plot the probability that snapshots of a given model exhibit this flip.  Unlike the signature in total intensity, this signature in linear polarization is stronger for MAD models than for SANEs, since SANEs are affected more strongly by Faraday depolarization \citep[e.g.,][]{Ricarte+2020,EHT8}.  In fact, of the SANEs, only the retrograde $R_\mathrm{high}=1$ models exhibit an authentic signal, originating from the flip in handedness at small radius also observable in total intensity.  Some prograde SANE $R_\mathrm{high}=10$ models appear as false positives only because $\angle \beta_2(\rho)$ fluctuates around 0.

\section{Discussion and Conclusions}
\label{sec:conclusion}

Within a Kerr BH's ergosphere, all material must co-rotate in an extreme case of frame-dragging.  In a retrograde system, infalling streams must transition from counter-rotating to co-rotating at some radius, leading to observable signatures in both total intensity and polarimetric images.  In total intensity, inflowing streams switch handedness due to this switch in tangential velocity.  In linear polarization, the handedness of linear polarization ticks also switches due to magnetic flux freezing into the inflowing plasma.  These signatures require observably emitting streams in the midplane, and thus significant jet emission may obfuscate this signature.

Our models assume a two temperature plasma consisting of ions and electrons in thermal distributions, but a non-thermal electron population is also expected to be present \citep[e.g.,][]{Ozel+2000,Ball+2018}.  Fortunately, since our signatures originate directly from the gas dynamics and magnetic field geometry, we do not expect them to be too sensitive to the details of the eDF.  Although we do not generate such an expansive library for other GRMHD models, we have also verified that our signatures appear in images generated by other GRMHD codes \citep{Chatterjee+2021,EHT5}.

\begin{figure*}
    \centering
    \includegraphics[width=0.99\textwidth]{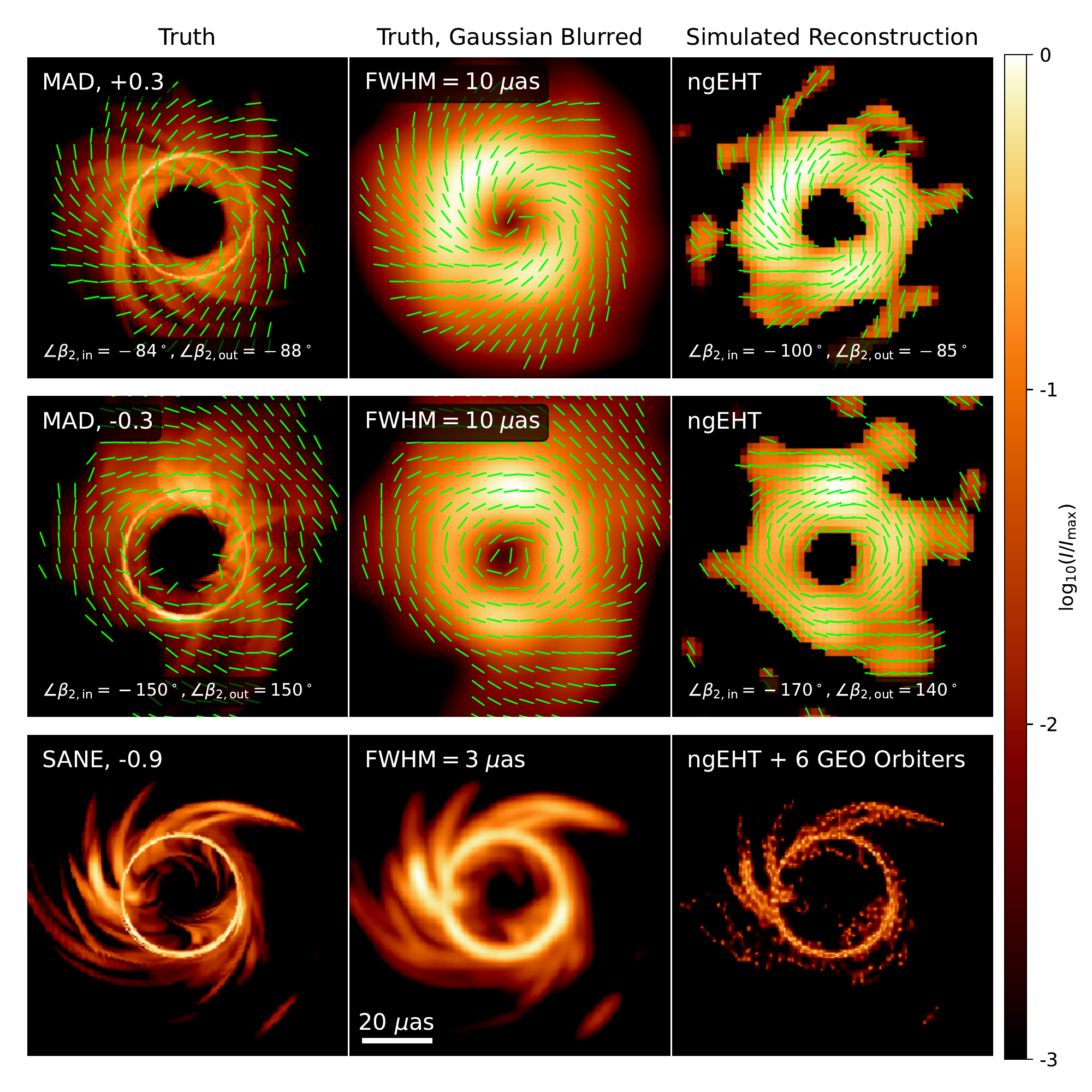}
    \caption{Example model images and their simulated reconstructions using ngEHT and potential space-based extensions.  Models are 345 GHz images corresponding to a MAD $a_\bullet=0.3$ $R_\mathrm{high}=10$ snapshot, a MAD $a_\bullet=-0.3$ $R_\mathrm{high}=10$ snapshot, and a SANE $a_\bullet=-0.9$ $R_\mathrm{high}=40$ snapshot respectively.  Perfect resolution images are shown in the left columns, images blurred with a Gaussian kernel are shown in the central columns, and simulated reconstructions are shown in the right columns, demonstrating that future arrays will be capable of observing these features.  The ngEHT would be capable of observing the flip in sign of $\angle \beta_2$ that occurs as a function of radius for retrogrades, while space-based extensions may be necessary to resolve the turnaround of individual streams.}
    \label{fig:pol_imaging}
\end{figure*}

In practice, observing these signatures will require imaging with significantly higher spatial resolution and dynamic range than the present EHT.  This motivates augmenting the EHT with higher observing frequencies, longer baselines, and additional stations.  In \autoref{fig:pol_imaging}, we perform simulated image reconstruction of three model images using two different futuristic arrays.  The first two rows correspond to simulated polarimetry using the ngEHT, while the third row corresponds to simulated imaging augmenting the ngEHT with 6 geosynchronous satellites.  Details of our simulated arrays and image reconstructions can be found in Appendix \ref{sec:simulated_imaging}.  All models are computed at 345 GHz and from top to bottom correspond to a prograde MAD, a retrograde MAD, and a retrograde SANE.  With polarimetry, ngEHT can detect the flip in the sign of $\angle \beta_2$ as a function of radius that occurs in our retrograde example.  Increasing the spatial resolution with space VLBI allows the array to resolve thin streams and directly observe their turnaround in total intensity.  Upon time-averaging, these thin streams are washed out, but the sign flip in $\angle \beta_2$ may persist, depending on the model, as we explore in \autoref{sec:time_averaging}.

The observational signatures described in this work provide novel tests of frame dragging in the strong field regime that will be accessible to extensions of the EHT.  Constraining the relative alignment of a BH's angular momentum and that of its accretion disk also has implications for feedback and spin evolution.  Prograde disks have larger jet efficiencies than retrograde disks \citep{Tchekhovskoy&McKinney2012,Narayan+2022}, and cosmic spin evolution depends on this relative alignment \citep[e.g.,][]{Volonteri+2013}.  In general, it is expected that the accretion disks of these systems will not be perfectly aligned or anti-aligned, but rather at a random angle with respect to the spin axis.  In such models, Lense-Thirring precession leads to a more rich dynamical system \citep[e.g.,][]{Fragile+2007,Chatterjee+2020,Liska+2021}.  How the signatures described in this work change in the general case remains an open and interesting question.  Among $R_\mathrm{low}=1$ models, MAD $a_\bullet=-0.5$ models in fact performed the best at passing simultaneous polarimetric constraints of M87* \citep{EHT8}.  If either Sgr A* or M87* has a retrograde disk, ngEHT could provide the first observations of frame dragging in the strong field regime.

\section{Acknowledgments}

We thank Sara Issaoun for a helpful script to perform simulated image reconstruction, Andrew Chael for help interpreting GRMHD dump files, Richard Qiu for a helpful script to solve for mass scaling, and Dominic Chang for insightful discussions about frame dragging.  This material is based upon work supported by the National Science Foundation under Grant No. OISE-1743747. Daniel C. M. Palumbo was supported by NSF grant AST-1716536 and the Gordon and Betty Moore Foundation grant GBMF-5278. Freek Roelofs was supported by NSF grants AST-1935980 and AST-2034306. This work was supported by the Black Hole Initiative at Harvard University, made possible through the support of grants from the Gordon and Betty Moore Foundation and the John Templeton Foundation. The opinions expressed in this publication are those of the author(s) and do not necessarily reflect the views of the Moore or Templeton Foundations.  Razieh Emami acknowledges the support by the Institute for Theory and Computation at the Center for Astrophysics as well as grant numbers 21-atp21-0077, NSF AST-1816420 and HST-GO-16173.001-A for very generous supports.  

\software{{\sc koral} \citep{Sadowski+2013,Sadowski+2014}, {\sc ipole} \citep{Moscibrodzka&Gammie2018}, Matplotlib \citep{matplotlib}, SciPy \citep{scipy}, NumPy \citep{numpy}, {\tt eht-imaging} \citep{ehtim}}

\bibliography{ms}

\appendix

\section{Generation of Model Images}
\label{sec:image_generation}

We use the software {\sc ipole} to perform polarized GRRT \citep{Moscibrodzka&Gammie2018}, using the standard imaging pipeline described in \citet{Wong+2022}.  In {\sc ipole}, first the null geodesic equation is solved from the camera through the source to determine the trajectories of photons which appear on the image.  Then, the equations of polarized radiative transfer are integrated forwards along the geodesic, using radiative transfer coefficients assuming thermal electron distribution functions \citep{Dexter2016}.  During GRRT, the free parameter $R_\mathrm{high}$ is introduced to scale the electron temperature relative to the ions via the prescription of \citet{Moscibrodzka+2016},

\begin{equation}
    \frac{T_i}{T_e} = R_\mathrm{high}\frac{\beta^2}{1+\beta^2} + \frac{1}{1+\beta^2},
\end{equation}

\noindent where $T_i$ and $T_e$ are the ion and electron temperatures respectively, and $\beta$ is the ratio of gas to magnetic pressure.  Such a prescription is physically motivated because the mean free path of particles in the plasma is much larger than the size of the system, and the ions are thought to receive more heating than electrons in large $\beta$ regions \citep{Rees+1982,Narayan&Yi1995,Quataert&Gruzinov1999,Howes2010,Kawazura+2019,Chael+2019,Mizuno+2021}.  Typically, $\beta$ is larger in the disk midplane, and smaller in the jet funnel.  SANE images tend to be more sensitive to $R_\mathrm{high}$ than MAD images, since they intrinsically have larger $\beta$ due to their weaker magnetic fields \citep{EHT5,Wong+2022}.

When producing images, we use a black hole mass of $6.2\times 10^9 \ M_\odot$, a distance of 16.9 Mpc \citep{Gebhardt+2011}, a field of view of 160 $\mu$as for most frequencies, and a pixel size of 0.5 $\mu$as, values chosen for consistency with previous EHT studies \citep{EHT5,EHT8}.  Since images become larger at 86 GHz, we adopt a field of view of 320 $\mu$as for this frequency.  We zero out the density in regions where plasma $\sigma > 1$, where numerical floors may artificially inject material.  Radiative transfer is integrated out to a radius of $100 \; GM_\bullet/c^2$.  Although polarized images can be sensitive to this outer integration radius due to Faraday rotation at large distances, this effect should be small for models of M87*, which we are viewing through an evacuated funnel region \citep{Ricarte+2020,EHT8}.

The equations of GRMHD are invariant under the transformation $u \to \mathcal{M}u$, $\rho \to \mathcal{M}\rho$, and $B \to \sqrt{\mathcal{M}}B$, where $u$ is the internal energy, $\rho$ is the mass density, $B$ is the magnetic field strength, and $\mathcal{M}$ is a scalar.  Hence, we have the freedom to select $\mathcal{M}$ to produce the correct flux for M87*, 0.5 Jy.  Since these simulations are run for so long, a constant value of $\mathcal{M}$ could lead to enormous departures from the target flux, due to the draining of the torus, for example.  Hence, we fit for scalars $a$ and $b$ and set $\mathcal{M} = \exp(a+bt)$, where $t$ is in units of $GM_\bullet/c^3$.  This allows us to remove secular trends in the flux due to slow changes in the accretion rate while preserving short-term variability.

Our image library includes 2 magnetic field states (MAD and SANE), 9 spins ($a_\bullet \in \{0,\pm0.3,\pm0.5,\pm0.7,\pm0.9\}$), and 4 electron temperature prescriptions ($R_\mathrm{high} \in \{1,10,40,160\}$) each imaged at 201 times (evenly spaced between a starting time of $10^4 \ GM_\bullet/c^3$ and an ending time of $10^5 \ GM_\bullet/c^3$ for MADs and $3\times 10^4 \ GM_\bullet/c^3$ for SANEs), at 4 frequencies $\nu \; [\mathrm{GHz}] \in \{86,228,345,690\}$.  This results in a total of 57888 images.

\begin{figure*}
    \centering
    \includegraphics[width=\textwidth]{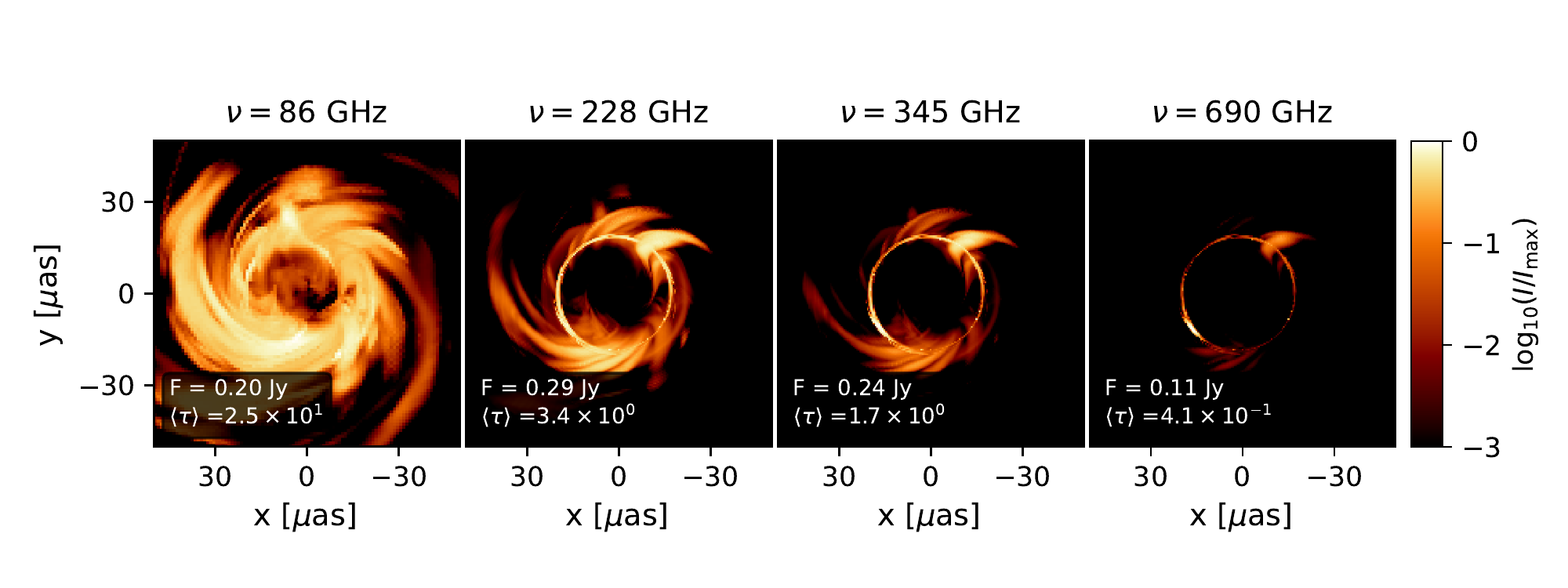}
    \caption{An example model from our library (SANE, $a_\bullet=-0.7$, $R_\mathrm{high}=40$) at the four frequencies included in our library, as indicated at the top of each panel.  Total flux density and image-averaged optical depth are written at the bottom left.  At 86 GHz, the image becomes larger and more optically thick than 228 GHz.  At 690 GHz, the spiral features probed in this work grow fainter relative to the photon ring and more difficult to observe.}
    \label{fig:frequency_gallery}
\end{figure*}

We plot an example snapshot from our library at these four different frequencies in \autoref{fig:frequency_gallery}, SANE, $a_\bullet=-0.7$, $R_\mathrm{high}=40$.  At the bottom left of each panel, we write the total flux density and an intensity-weighted average of the optical depth across all pixels in the image.  Compared to 228 GHz, the 86 GHz image is larger, thus probing larger radii, but more optically thick, potentially obscuring signals at the center of the image.  At 690 GHz, the spiral features become faint relative to the photon ring and more difficult to observe.  Typically, although these general trends hold, our MAD models evolve less dramatically with frequency, since their spectral indices are shallower \citep{Ricarte+2022}.  Fortuitously, 228-345 GHz is typically the ideal frequency at which to observe the retrograde signatures described in this work.

\section{Fourier Decomposition into Logarithmic Spirals}
\label{sec:fourier_decomposition}

\begin{figure*}
    \centering
    \includegraphics[width=\textwidth]{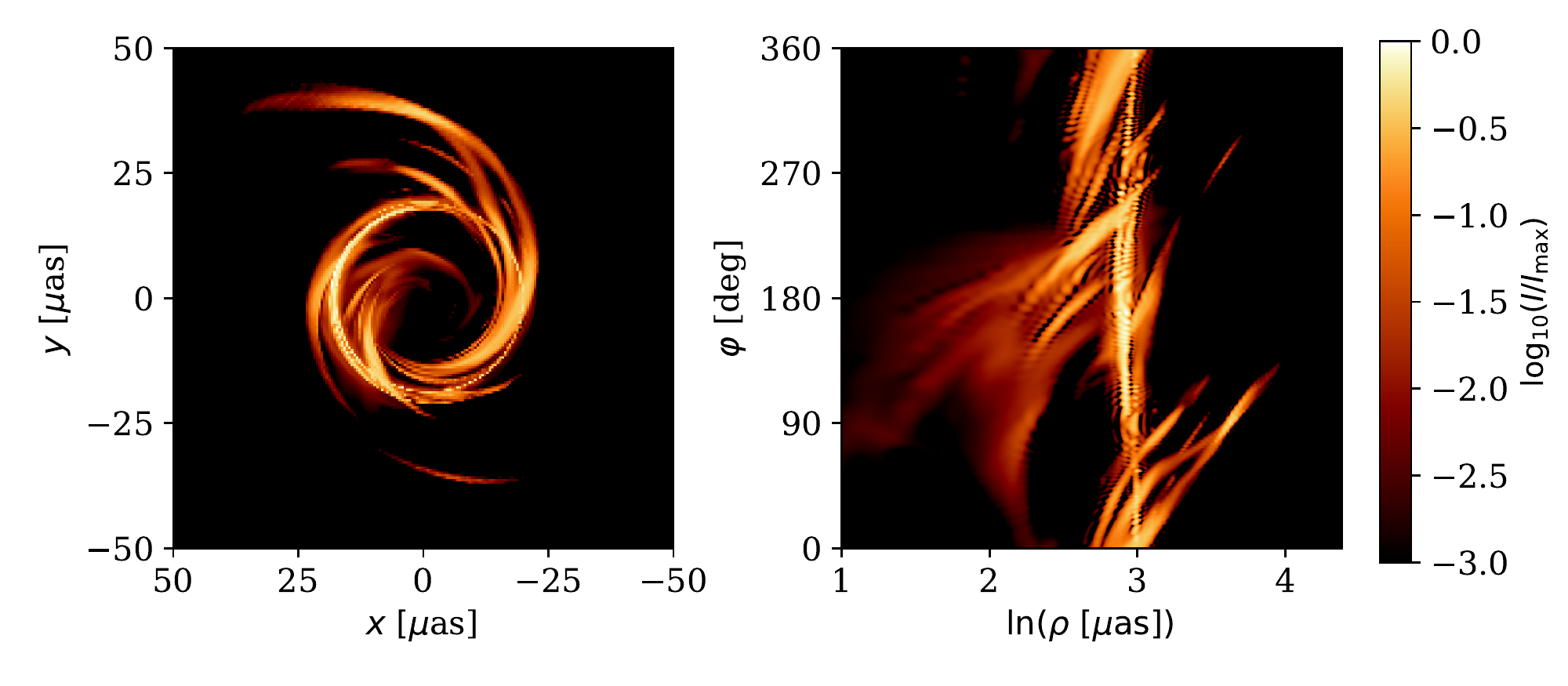}
    \caption{An example image from our library (SANE, $a_\bullet=+0.5$, $R_\mathrm{high}=160$, 228 GHz) and its ``unwrapped'' equivalent.  Plotted in $(\ln\rho, \varphi)$ coordinates, logarithmic spirals with different pitch angles appear as diagonal lines with different slopes.  Meanwhile, the circular photon ring becomes a nearly vertical line.}
    \label{fig:unraveled_example}
\end{figure*}

There has been an extensive history of decomposing images into logarithmic spirals in the galaxy morphology community, and we follow much of the formalism developed in these previous works \citep[e.g.,][]{Kalnajs1975,Considere&Athanassoula1982,Puerari+1992,Puerari+2000,Barbera+2004,Davis+2012}.  The basis function for our decomposition is the logarithmic spiral, which can be expressed in polar coordinates as 

\begin{equation}
    \rho = \rho_0 e^{\varphi \tan\Phi}, 
    \label{eqn:basis_function}
\end{equation}

\noindent where $\rho$ is the image radius, $\rho_0$ is the initial radius at $\varphi=0$, $\varphi$ is the angular coordinate in the image, and $\Phi$ is the pitch angle of the spiral.  When plotted ``unwrapped'' via interpolation onto $(\ln\rho,\varphi)$ coordinates, logarithmic spiral images appear as straight lines.  We show one example snapshot from our library (SANE, $a_\bullet=+0.5$, $R_\mathrm{high}=160$, 228 GHz) and its unwrapped equivalent in \autoref{fig:unraveled_example}.  It is characterized by logarithmic spirals with different pitch angles, and thus its unwrapped image features diagonal lines with different slopes.  The photon ring is mapped to a nearly vertical line.

For an annulus in the image defined by $\rho_\mathrm{min} < \rho < \rho_\mathrm{max}$, the amplitude of each Fourier component of the image is given by 

\begin{equation}
    A(p,m) = \frac{1}{D} \int_{-\pi}^{+\pi}\int_{\ln \rho_\mathrm{min}}^{\ln \rho_\mathrm{max}} \frac{dF}{du d\varphi}(\rho,\varphi) e^{-i(m\varphi + pu)}dud\varphi,
\end{equation}

\noindent $dF/dud\varphi$ is the distribution one is decomposing, $u\equiv\ln\rho$, $m$ is the number of spiral arms or harmonic modes, $D$ is a normalization factor given by

\begin{equation}
    D = \int_{-\pi}^{+\pi}\int_{\ln \rho_\mathrm{min}}^{\ln \rho_\mathrm{max}} \frac{dF}{du d\varphi}(\rho,\varphi) du d\varphi,
\end{equation}

\noindent and $p$ is the variable associated with the pitch angle $\Phi$ via

\begin{equation}
    \tan(\Phi) = -\frac{m}{p}.
    \label{eqn:p_to_Phi}
\end{equation}

\noindent From \autoref{eqn:p_to_Phi}, note that clockwise modes (defined moving from the origin outward) are associated with negative pitch angles and therefore $p>0$, while counterclockwise modes are associated with positive pitch angles and $p<0$.  Although it is not used in our analysis, one can also derive the phase $\Psi$ of a spiral mode via

\begin{equation}
    \tan\Psi = \frac{\mathrm{Im}(A(p,m))}{\mathrm{Re}(A(p,m))},
\end{equation}

\noindent which describes how the spiral is oriented.  Note that $\rho_0$ in \autoref{eqn:basis_function} is determined by $\Psi$ via $\rho = \rho_0e^{\varphi \tan\Phi} = e^{\varphi \tan\Phi+\ln\rho_0} = e^{\varphi \tan\Phi+\Psi}$.

In practice, we produce model images $dF/dxdy(x,y)$ with units of intensity ($\mathrm{Jy} \; \mathrm{\mu as}^{-2}$) and even sampling in $x$ and $y$.  Thus, it is more convenient to change variables and compute

\begin{equation}
    A(p,m) = \frac{1}{D} \iint \frac{1}{\rho^2} \frac{dF}{dxdy}(x,y) e^{-i(m\varphi + p u)} \zeta(\rho) dxdy,
\end{equation}

\noindent and likewise for $D$, where $1/\rho^2$ is the Jacobian associated with this change of variables, and $\zeta(\rho)$ is a masking function, equal to 1 if $\rho \in (\rho_\mathrm{min},\rho_\mathrm{max})$ and 0 otherwise.  Due to the large dynamic range in intensity spanned by these images, we perform these decompositions not on the raw image, but rather on the logarithm of the image, with an explicitly specified dynamic range of 3 orders of magnitude.  This mitigates a potential bias in our decomposition towards smaller image radii that could occur due to the rapid falloff of flux with radius.  To do so, we take

\begin{align*}
    \frac{dF}{dxdy} = \begin{cases}
    \log_{10}(I/I_\mathrm{max}) + 3 &\text{\ if $\log_{10}(I/I_\mathrm{max}) \geq -3$}\\
    0 &\text{\ if $\log_{10}(I/I_\mathrm{max}) < -3$}
    \end{cases}
\end{align*}

\noindent where $I$ is the intensity in each pixel.  Finally, we also recenter images on the centroid of the inner shadow, the lensed image of the equatorial horizon, which is a natural origin for these spiral features.  We calculate the centroid location using the fitting formula as a function of spin and inclination provided in \citet{Chael+2021}.  In practice it is quite close to the image origin for models of M87* due to its low inclination, but not close to the centroid of the photon ring for high spin.

When performing these decompositions, we consider $m \in \{0,1,2,...,20\}$ and $p \in [-100,100]$ with $\Delta p = 0.02$.  We comment that $m \gtrsim 5$ is necessary to capture the thin streams of material plunging into the black hole, and $\Delta p \lesssim 0.1$ is sometimes necessary to resolve narrow peaks in $A(p,m)$ as a function of $p$ and produce inverse Fourier transforms without clear numerical artifacts.  In practice, when applying this technique to real data, we expect that we would set the maximum value of $m$ to a value that captures the main image features but avoids reconstructing image artifacts, depending on the dataset.  Since we are interested in searching for a switch in handedness as a function of radius, we compute these decompositions separately for $\rho \in [1\; \mu\mathrm{as},15\;\mu\mathrm{as}]$ and $\rho \in [31\; \mu\mathrm{as}, 50\;\mu\mathrm{as}]$.  These two regions deliberately avoid the photon ring, which for M87* appears at $\rho \approx 19 \; \mu\mathrm{as}$ \citep[e.g.,][]{EHT5}.  We find that our results are insensitive to the exact values of the radii defining the outer annulus, as long as it is outside the photon ring and within a radius at which the image produces an appreciable amount of flux.

Once the Fourier coefficients are computed, it is illustrative to compute the inverse Fourier transform (IFT), which is given by \citep{Seigar+2005}

\begin{equation}
    S(u,\varphi) = \sum_m S_m(u) e^{i m \varphi} \label{eqn:inverse}
\end{equation}

\noindent where

\begin{equation}
    S_m(u) = \frac{D}{e^{2u}4\pi^2}\int A(p,m) e^{i p u} dp. \label{eqn:inverse_integral}
\end{equation}

\noindent This methodology is used to construct the IFT images shown in \autoref{fig:examples}.

\section{Simulated Image Reconstructions}
\label{sec:simulated_imaging}

The next-generation Event Horizon Telescope (ngEHT) will be a significant enhancement of the EHT, with $\sim 10$ new stations added, a quadrupled bandwidth, and the possibility of observing at frequencies up to 345 GHz \citep{Doeleman+2019}. The expanded array has improved short and medium baseline coverage, vastly improving sensitivity to diffuse structures over EHT data. Thus, we might expect ngEHT images of M87* to be sensitive to the diffuse, low surface brightness signatures described in this work. 

In this work, we simulate single-day 345 GHz observations of M87* with an ngEHT reference array. This array consists of the EHT array in 2022 plus ten additional sites, and was used for the ngEHT Analysis Challenges \citep{Raymond+2021, Doeleman+2022, Roelofs+2022}\footnote{See \url{https://challenge.ngeht.org/challenge1/} for more details on data generation.}.  For the ground stations, we add thermal noise corresponding to a system temperature of 100 K, aperture efficiency of 0.42, and, for new ngEHT stations, a dish diameter of 6 m. Data were assumed to be amplitude calibrated but not phase calibrated.

The ngEHT images in \autoref{fig:pol_imaging} were reconstructed in a two-step imaging process that first produced a Stokes $I$ image with amplitude and closure phase information, and then produced a polarized image using fractional polarized visibility $\Breve{m}$. The Stokes $I$ reconstruction weakly regularized entropy, sparsity, and total variation (see definitions in \citet{EHT4}). The polarimetric reconstruction used mild regularization of polarized entropy and total variation \citep{Holdaway_1990,Chael_2016}.

Since the angular resolution attainable with ground-based VLBI is limited by the size of the Earth and severe atmospheric absorption and turbulence towards higher submillimeter frequencies, the natural next step will be to employ a Space VLBI array. Several submillimeter Space VLBI array concepts aiming at high-resolution observations of black hole shadows have been proposed \citep[e.g.][]{Palumbo+2019, Roelofs+2019, Pesce+2019, Fish+2020, Johnson+2020, Andrianov+2021}, with antennas in orbits ranging from Low Earth Orbits to the Sun-Earth L2 Lagrange point. 

In addition to the synthetic ngEHT data, we simulate 345 GHz observations with the same ngEHT reference array plus six antennas in geosynchronous orbits (GEO). In order to sample a wide range of baseline length and orientations over a one-day observation, we place three satellites in equatorial (geostationary) orbits, and three satellites in the geosynchronous orbit normal to the line of sight to M87. For the stations in geosynchronous orbits, we assumed a system-equivalent flux density (SEFD) of 10000 Jy to set the magnitude of the thermal visibility noise. The resulting maximum baseline length is $\sim 100$ G$\lambda$, corresponding to an angular resolution of $\sim 2$ $\mu$as.

The ngEHT+GEO image in \autoref{fig:pol_imaging} (lower right) was reconstructed with $\texttt{eht-imaging}$ using visibility amplitudes and closure phases \citep{Chael+2018}, and no regularization apart from total flux and image centroid regularization. Since the aim of our Space VLBI imaging simulations is to provide a proof of concept for the observability of the spiral handedness flip in retrograde models, we did not optimize the Space VLBI array configuration, observing strategy, or image reconstruction process for this work. Further study will be needed to set the minimum system requirements for observing the spiral handedness flip.

\section{Effect of Time-Averaging}
\label{sec:time_averaging}

\begin{figure*}
    \centering
    \includegraphics[width=\textwidth]{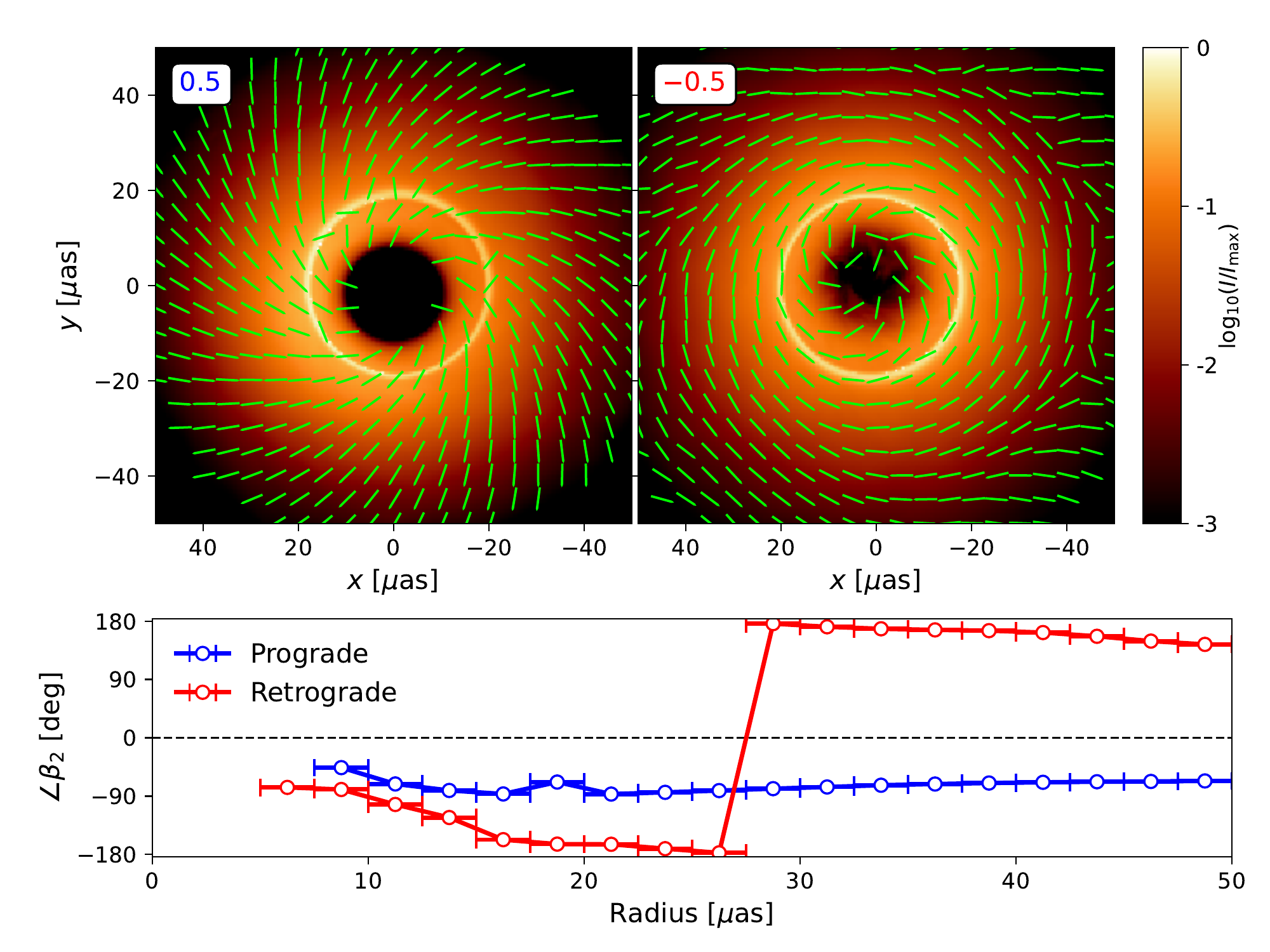}
    \caption{Time-averaged images for a MAD $a_\bullet=0.5$ model on the left and $a_\bullet=-0.5$ on the right, each with $R_\mathrm{high}=1$ at 228 GHz.  In the lower panel, we plot $\angle\beta_2$ as a function of radius for these models, where the blue curve corresponds to the prograde and the red curve corresponds to the retrograde.  In this model, time-averaging washes away the individual ``S''-shaped streams, but preserves the flip in $\angle\beta_2$ as a function of radius in the retrograde.}
    \label{fig:time_average}
\end{figure*}

Here, we briefly discuss whether the signatures described in this work can survive under time-averaging.  This is most relevant for observations of Sgr A*, which varies on timescales shorter than a night of observations.  In \autoref{fig:time_average}, we plot the unblurred time-averaged image of two MAD models, with spin $a_\bullet=0.5$ on the left and $a_\bullet=-0.5$ on the right.  When time-averaged, the transient spiral arm features of all models are washed out and can no longer be observed.  However, the linear polarization signature may still be accessible in some models.  In the bottom panel, we plot $\angle \beta_2$ as a function of radius, centered on the centroid of the inner shadow.  Only annuli contributing at least one one-thousandth of the total flux are included.  While the prograde model exhibits negative $\angle\beta_2$ at all radii, the retrograde model flips sign, as expected.

\end{document}